\documentclass[letter,traditabstract]{aa}
\usepackage{natbib}
\bibpunct{(}{)}{;}{a}{}{,} 
\usepackage{graphicx}
\usepackage{txfonts}
\newcommand{\corr}{\rm}
\begin{document}
\title{Multi-frequency VLBA study of the blazar S5 \object{0716+714} during the active state in 2004}
%
\subtitle{I. Inner jet kinematics}
\author{E. A. Rastorgueva \inst{1} 
  \and
  K.~Wiik\inst{1}
  \and
  T.~Savolainen\inst{2}
  \and
  L.~O.~Takalo\inst{1}
  \and
  E.~Valtaoja\inst{1}
  \and
  Y.~N.~Vetukhnovskaya\inst{3}
  \and
  K.~V.~Sokolovsky\inst{2, 4}
}
\offprints{E. Rastorgueva, \email{eliras@utu.fi}}
\institute{Tuorla Observatory, Dept. of Physics and Astronomy, University of Turku, \\
  V{\"a}is{\"a}l{\"a}ntie 20, 21500 Piikki{\"o}, Finland;
  \and
  Max-Planck-Institut f{\"u}r Radioastronomie, \\
  Auf dem H{\"u}gel 69, 53121 Bonn, Germany;
  \and
  Pushchino Radio Astronomy Observatory,  Lebedev Physical Institute, \\
  Leninskii Prosp. 53, 119991 Moscow, Russia;
  \and
Astro Space Center of Lebedev Physical Institute,
Profsoyuznaya 84/32, 117997 Moscow, Russia
}
\date{Received / Accepted } \abstract{We observed the blazar \object{0716+714} with the
  VLBA during its active state in 2003-2004. In this paper we discuss
  multi-frequency analysis of the inner jet (first 1~mas) kinematics. The
  unprecedentedly dense time sampling allows us to trace jet components
  without misidentification and to calculate the component speeds with good
  accuracy. In the smooth superluminal jet we were able to identify and track
  three components over time moving outwards with relatively high apparent
  superluminal speeds (8.5-19.4\,$c$), which contradicts the
  hypothesis of a stationary oscillating jet in this source. Component
  ejections occur at a relatively high rate (once in two months), and they are
  accompanied by mm-continuum outbursts. Superluminal jet components move
  along wiggling trajectories, which is an indication of actual helical
  motion. Fast proper motion and rapid decay of the components suggest that
  this source should be observed with the VLBI at a rate of at least once in
  one or two months in order to trace superluminal jet components without
  confusion.}  
\maketitle

\section{Introduction}
The blazar S5 \object{0716+714} is one of the most active sources of its
class. It is highly variable on time scales from hours to months
at all observed wavelengths from radio to $\gamma$-rays. Intra-day
variability (IDV) was detected in the source in the optical,
millimetre, and radio bands \citep[e.g.][and references
therein]{montagni2006, agudo2006}. The IDV in the different bands
of electromagnetic radiation is strongly correlated
\citep{quirrenbach1991, wagner1996, stalin2006, fuhrmann2008},
which suggests its intrinsic origin. Redshift of the source is
0.31, as recently estimated by \citet{Nilsson_z2008}, who observed
the host galaxy in the I-band during the quiescent state of the
active nucleus.

Very long baseline interferometry studies (VLBI) show a
core-dominated jet pointing to the north \citep{jorstad2001,
bach2005}, and VLA data show a halo-like jet misaligned with it by
$\sim$~90$^{\circ}$ \citep{wagner1996}. There are several
scenarios for the milliarcsecond scale jet kinematics. The most
recent studies propose two very different interpretations for the
long time series ($\sim$10~yr) of centimetre VLBI observations.
One scenario consists of fast superluminal components moving along
the jet with speeds ranging from 5\,$c$ to 16\,$c$
\citep{jorstad2001,
  bach2005}. An alternative interpretation suggests that instead of travelling
along the jet, components are ``oscillating'' around their mean
position due to the jet precession. Component speeds, calculated
taking position angle changes into account, range from 5\,$c$ to
10\,$c$ \citep{britzen2006}.

This paper is the first one in a series of papers that describe
changes in the parsec-scale structure of \object{0716+714} during the
active period in 2003-2004. Here we present the kinematics of the
inner jet of \object{0716+714}. In Paper II we will discuss the suitability
of different data imaging methods for restoration of the source
structure in the case of a very smooth brightness distribution
along the jet. Source polarisation, quantitative analysis of the
flux density evolution and spectral properties will be published
in the subsequent papers. Here we use a standard cosmological
model with a flat universe and values of Hubble constant
$H_{o}$=71\,km~s$^{-1}$Mpc$^{-1}$, energy density of matter
$\Omega_{m}$=0.3, and cosmological constant energy density
$\Omega_{\Lambda}$=0.7.

\section{Observations and data analysis}

\begin{figure*}
   \centering
   \includegraphics[angle=-90,width=0.85\textwidth]{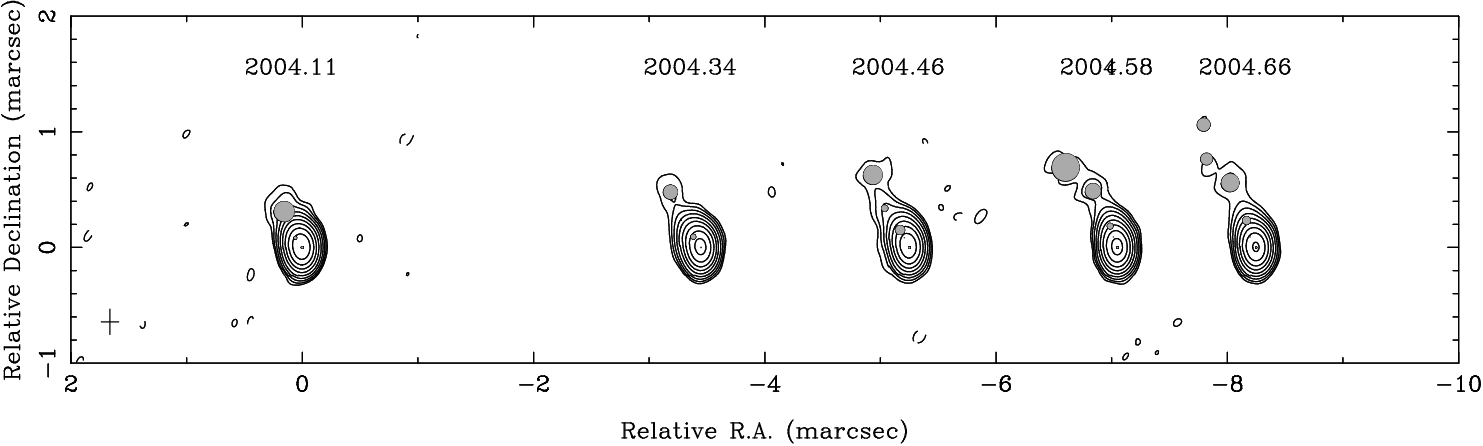}
   \caption{Contour maps of \object{0716+714} at 43~GHz for five epochs convolved with
     the same beam. Contour levels are 7, 14, 28, 56, 112, 224, 448, 896, 1792
     mJy/beam. The distance between maps in horizontal direction is
     proportional to the time separation between epochs. Grey circles
     represent Gaussian model components.}
    \label{43maps}
\end{figure*}

\object{0716+714} was observed with the Very Long Baseline Array (VLBA).
Five observations, separated by approximately a month (2004, Feb
10 (A), May 03 (B), Jun 18 (C), Jul 29 (D) and Aug 29 (E)), cover
a time period of the source's active state. This is an
unprecedentedly high time resolution for this source. Each
observation lasted for nine hours and utilised five frequencies
(86, 43, 22, 5 and 1.6 GHz) in dual polarisation mode. Data were
correlated at the Socorro VLBA correlator. {\it A priori}
amplitude and phase calibration was performed at Tuorla
Observatory using standard procedures of the AIPS package. Imaging
was done using Caltech Difmap package employing CLEAN algorithm.

\subsection{Model fitting and error analysis}

After the final self-calibrated data and CLEAN image were obtained, we fitted a source model consisting of Gaussian components to the data in the
$(u,v)$-plane using Difmap procedure Modelfit. We used only circular Gaussian
components in order to reduce the number of free parameters, and we tried to
use as few components as possible. For all models we took the brightest
component at the end of the jet as a reference (the ``core'') and the
component positions were measured relative to it.

The errors in position and flux density of the components were
calculated using Difwrap package \citep{lovell2000}, which employs
Modelfit in a loop for finding maximum possible deviation of the
component parameter from the best-fit value. We used it in a
procedure proposed by \citet{savolainen2006}, applying our own
criteria for limiting error determination, which will be described
in detail in the next paper.

Expected values of ejection epochs and corresponding errors were
determined using Monte Carlo simulation, as well as expected
values of component positions and errors in Ra-Dec coordinates.
Each component was replaced with a Gaussian distribution of
values, representing the errors in that point. A weighted
least-squares fit was calculated for each set in the case of
ejection epoch determination, producing a (non-Gaussian)
distribution of zero crossings. The expected value and standard
deviation were calculated from these distributions using standard
equations.

\subsection{Cross-identification of  the model components}

Model fitting of the multi-frequency VLBI data set simultaneously
is not a trivial task, because resolution, values of model
parameter errors, and position of the core all depend on the
frequency. Therefore, we have to treat different bands separately.
In this paper we discuss the kinematics of the inner jet mainly
using data obtained at 43~GHz. In addition to that, the data set
provides models of the same region at 86 and 22~GHz. Even though
the
  angular resolution at 22~GHz is only slightly lower ($\approx$0.25~mas) than
  at 43~GHz ($\approx$0.15~mas), this difference is important for the small
  inner jet region with fast moving components (see Sect.~\ref{superlum}).
Slightly higher resolution is provided by 86~GHz, but the data
quality at this frequency allows us to obtain 86~GHz images only for
two epochs of our experiment. Therefore, our kinematic model is based on 43 and 22~GHz data (with emphasis on 43~GHz), while 86~GHz data
is used for a consistency check of the models. All 22, 43 and 86~GHz models
will be used for the subsequent component spectra analysis.

\begin{figure}
  \centering
 \includegraphics[angle=-90,width=0.5\textwidth]{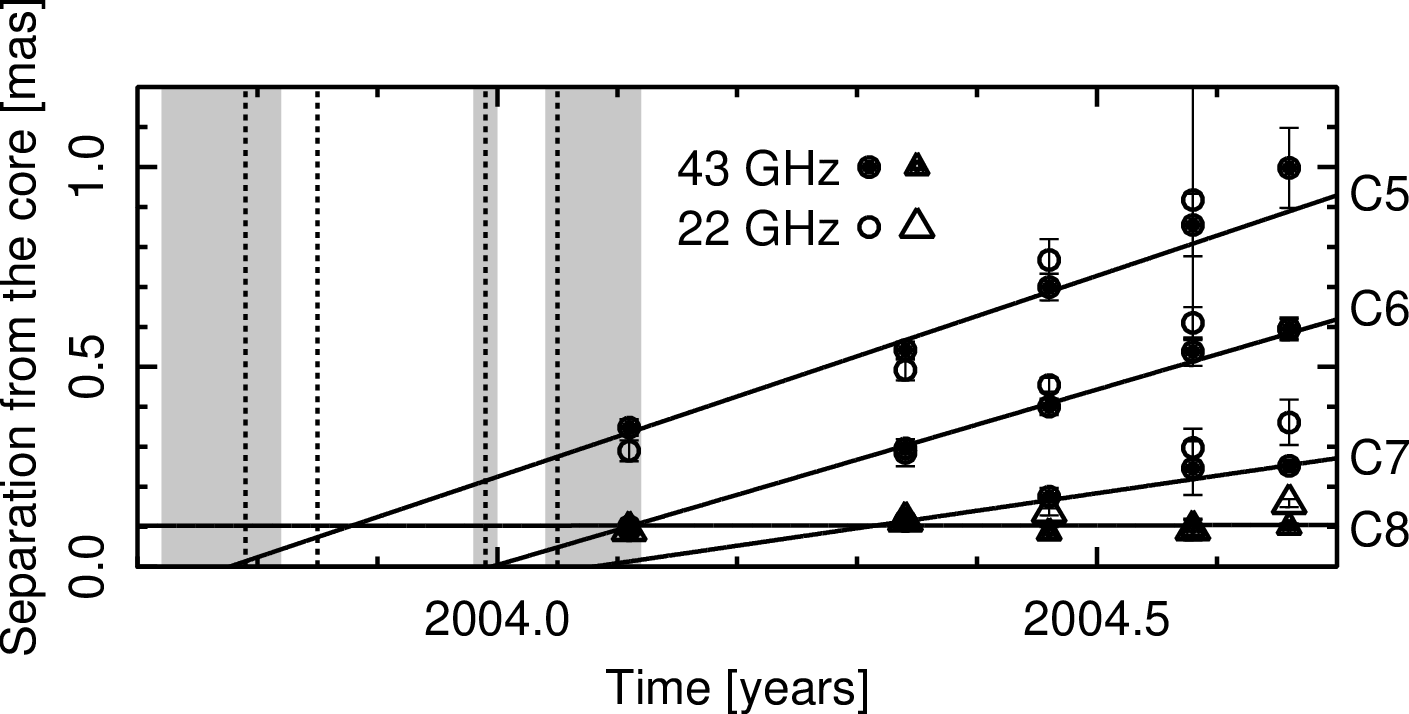}
 \caption{Radial distances from the core as a function of time with
   linear fits. Stationary component C8 is plotted with
   triangles. Shaded area represents a 1$\sigma$ error of component ejection
   time, vertical dashed lines mark the beginning of the mm flare.}
 \label{43models}
\end{figure}

Models provided by the Modelfit procedure are not a unique
representation of the data, they are only one of all the possible
ways to parametrise the image. Therefore, we had to work out tests
that help us to estimate the goodness of the model fit. We
consider a model acceptable when the following requirements are
fulfilled: a) model provides a good fit to the visibility data; b)
if convolved with the restoring beam, it provides a brightness
distribution similar to that of the final CLEAN map; c) all models
at the same frequency are consistent between epochs; i.e.,  most
of the components are traceable over time; and d) models at the
same epoch are consistent between frequencies.

Condition a) is fulfilled if the model-fitting process was carried
out correctly and verified by visual inspection of the
correspondence between the observed and model $(u,v)$-data.
Condition b) is checked by simple visual inspection of the
resulting images. Condition c) is verified by comparison of
component positions at different epochs and flux density
evolution, no unphysically rapid changes should be seen. In the
presence of outward motion, goodness of the linear fit is also a
good indicator. For condition d) we use the following rule:
components from two close frequencies with comparable resolution
should coincide within positional errors and have flux densities
that do not require unphysically steep spectral indices (or at
least have similar flux density-time dependence). Also, if two
components are resolved at higher frequency, while at the lower
frequency they appear as one component, the distance between those
components should not be greater than the resolution at the lower
frequency, and their total flux density should be close to the
flux density of the unresolved one. Our final model set seems to
meet all those requirements with good accuracy. Therefore, our
43~GHz kinematic model is self-consistent over time and is also in
good agreement with both neighbouring frequencies.

Cross-identification of the model components between the epochs is based
on their distance from the core, position on the (Ra,Dec)-plane and their flux
densities. In this paper, we rely on the accuracy of the {\it a priori}
amplitude calibration of the VLBA, which has been previously shown to
  achieve $\approx$5$\%$ at 43~GHz \citep{savolainen2008}. Distribution of
the polarised intensity along the jet is smooth and uniform, therefore it was
not possible to use polarisation information to cross-identify the
  components.

\section{Results}\label{superlum}\label{struct}
The VLBA maps of \object{0716+714} at 43 GHz are presented in
Fig.~\ref{43maps}. The date of each observation is indicated in
the plot.  Well-collimated one-sided inner jet of \object{0716+714} spans
up to 1~mas pointing to the position angle of $\sim$27$^{\circ}$
and has a smooth intensity distribution.
\begin{table}
\caption[]{Proper motions and apparent speeds of components.}
\label{app_motion} \centering
\begin{tabular}{cccc}
\hline\hline
Comp. & $\mu$, mas/yr & $\beta_{app}$, c & Ejection    \\
name  &               & (z=0.31)      & epoch, yr  \\
\hline
C5$^{43}$& 1.01$\pm$0.09 & 19.4$\pm$1.7 & 2003.77$\pm$0.05  \\
C5$^{22}$& 1.11$\pm$0.13 & 21.3$\pm$2.5 & 2003.86$\pm$0.05  \\
C6$^{43}$& 0.88$\pm$0.03 (1.04$\pm$0.07)$^{*}$ & 16.9$\pm$1.6 &  2003.99$\pm$0.01    \\
C6$^{22}$& 0.98$\pm$0.04 & 18.9$\pm$0.8  & 2004.02$\pm$0.01 \\
C7$^{43}$& 0.44$\pm$0.04 (0.44$\pm$0.16)$^{*}$ & 8.5$\pm$0.8 & 2004.08$\pm$0.04   \\
C7$^{22}$& 0.76$\pm$0.14 & 14.6$\pm$2.6  & 2004.18$\pm$0.04 \\
C8$^{43}$& 0.00$\pm$0.01 & -- & -- \\
C8$^{22}$& 0.12$\pm$0.02 & -- & -- \\
\hline
\end{tabular}
\begin{list}{}{}
\item[$^*$] In parenthesis: fit performed using travelled distance
\item[$^N$] Values obtained for 43 and 22 GHz are marked with corresponding number
\end{list}
\end{table}
A total of four components were detected during the observations,
and one of them (C8) is stationary. Their radial distances as a
function of time with error bars and linear fits for proper
motions are presented in Fig.~\ref{43models}. The plotted linear
fits were performed for 43~GHz data (filled triangles/circles). It
is clear that most 22~GHz components (open triangles/circles)
coincide with those at 43~GHz within one or two sigma. At epochs A
and B, components C6 and C7, respectively, are not resolved from
C8. Therefore we took positions of C8 as their positions at these
epochs. At epoch C, components C7 and C8 are not resolved at
22~GHz. Only one moving component (C5) was observed at all five
epochs. At epoch E its flux density faded, and it split into two
components of comparable flux densities. Their positions
  were averaged in order to perform the linear fit. Based on C5, we estimated
  the component lifetime (from ejection to fading away) to be ten
  months.

Proper motions, superluminal speeds, and ejection epochs for all
four components for 22 and 43~GHz are presented in
Table~\ref{app_motion}. Components C6 and C7 have substantially
curved trajectories, therefore, for them we present proper motions
and apparent speeds calculated using linear fit for both radial
and travelled distance. Errors in the radial distance fit
parameters were calculated using components' positional errors,
while travelled distance fit errors were only calculated from the
scatter. Therefore the former are substantially smaller. It is
clear that values of proper motions and apparent speeds obtained
by both methods agree within errors, which indicates that the
wiggling of the component trajectories (see below) does not
influence the determination of the component speeds.
\begin{figure}
 \centering
  \includegraphics[angle=0.0,width=0.30\textwidth]{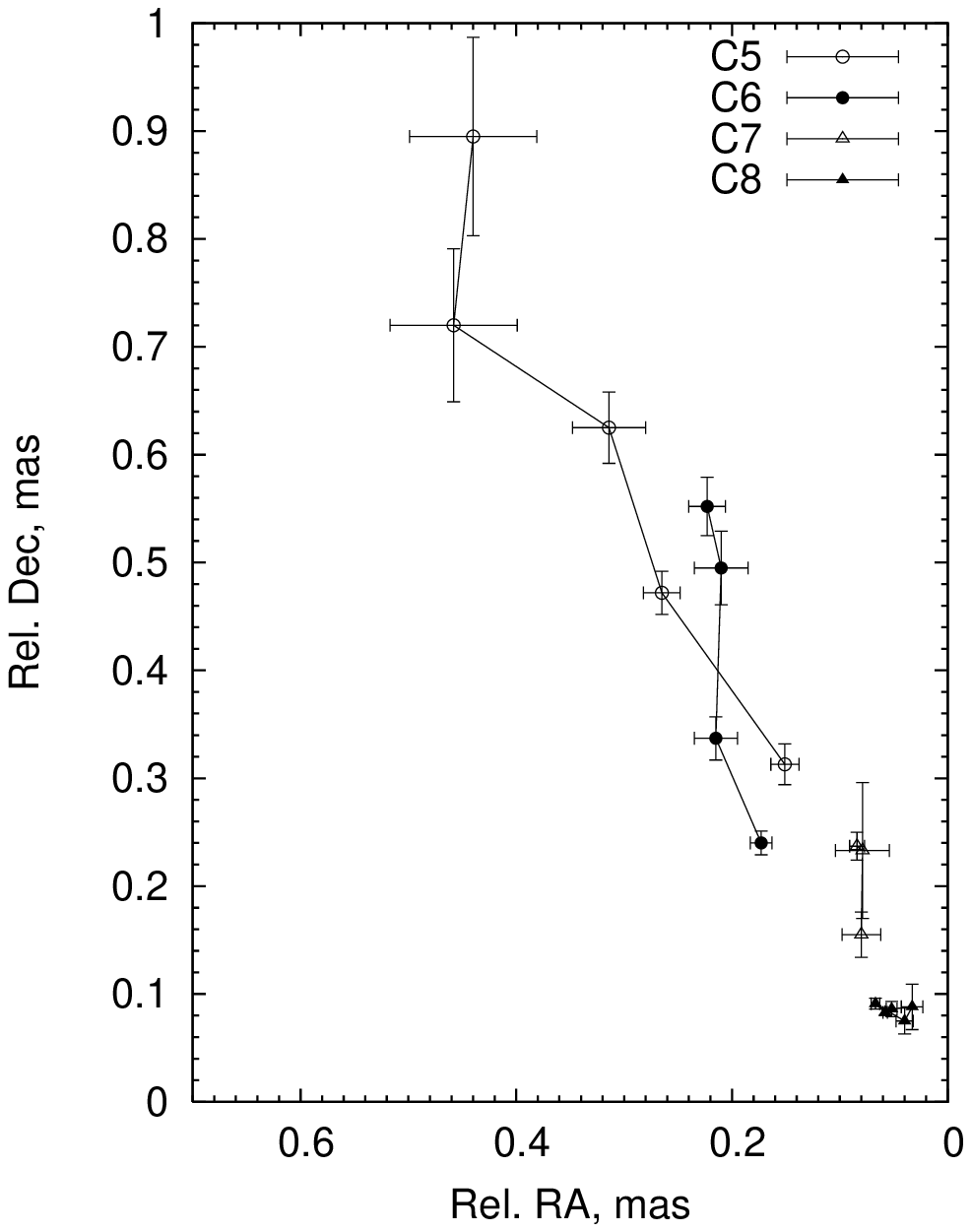}
  \caption{Trajectories of motion of the components identified at
    43~GHz.}
  \label{wiggl}
\end{figure}

The jet components have wiggling trajectories along the jet, with
the deviation from the overall jet direction greater than
positional errors (see Fig.~\ref{wiggl}). We have to notice that
this wiggling is less pronounced at lower frequencies (22~GHz),
which suggests that the inner jet is resolved in transverse
direction at 43~GHz.

\section{Discussion}
\subsection{Jet parameters}
Although kinematics results obtained for 22 and 43~GHz are similar, all
further calculations were performed for the apparent speeds obtained for
43~GHz, which seem to us more reliable due to the better resolution of
images. The estimated apparent speeds of the jet components are superluminal
with a maximum value of 19.4\,$c$. The viewing angle that maximises
the apparent speed is
$\theta_\mathrm{SL}=\arctan$($\beta_\mathrm{app}^{-1})=3.0^\circ$,
corresponding to a minimum Lorentz factor of $\gamma_\mathrm{SL}=\gamma_\mathrm{min}=(1+\beta_{app}^{2})^{1/2}=19.5$ and to a Doppler
  factor of $\delta_\mathrm{SL}\approx19.4$. If $\theta$ differs from
$\theta_\mathrm{SL}$, these values could be even higher; however,
19.4 is high enough to explain the brightness temperature,
estimated during the outbursts in the framework of the standard
synchrotron source model with a power-law electron energy
distribution \citep{ostorero2006}. Our results indicate slightly
higher flow speeds than previous VLBI kinematic studies by
\citet{bach2005} and agree with results of \citet{jorstad2001}.
However, radio variability gives $\delta_\mathrm{var}${\corr=10.9} \citep{hovatta2008a}. This value could be achieved for the
observed values of the apparent speed if the viewing angle
$\theta_\mathrm{var}=\arctan(2
\beta_\mathrm{app}/(\beta_\mathrm{app}^{2}+\delta_\mathrm{var}^{2}$-1){\corr=4.5$^\circ$}, which is lower
than the maximum value of 5.9$^\circ$, calculated according to the
relation $\sin\theta_\mathrm{max}=2\beta_\mathrm{app}/(1+\beta_\mathrm{app}^{2})$. Therefore, our
results are also consistent with calculations based on the source
variability.

\subsection{Correspondence of component ejections to mm flares}
According to \citet{savolainen2002}, an ejection of superluminal
components in blazars in most cases coincide with the beginning of
the mm-continuum flare. In September-October 2003 \object{0716+714}
underwent an outburst in mm and radio bands. The source had a
large flare with several peaks and a complicated structure in the
millimetre band (T. Hovatta and A. L{\"a}hteenm{\"a}ki, priv.
comm.). In the period under consideration, four flares were
registered at 37~GHz, starting at 2003.79, 2003.85, 2003.99, and
2004.05. As a flare start we took one e-folding time before the
maximum. Three of them coincide with the ejection times of
components C5, C6, and C7 (see Fig.~\ref{43models} and
Table~\ref{app_motion}).

\subsection{Wiggling trajectories}
We noticed that trajectories of the jet components are not
straight. The standard shock-in-jet model of AGN
\citep{marscher1996} suggests straight outward movement of the
emitting component along the jet, but several alternative models
state that emitting plasma clouds may move along helical paths.
Although we do not have enough data to distinguish between those
models, it might be interesting to apply them to this source in
the future.

A two-fluid model of the jet \citep{sol1989, despringre1997}
presents the jet itself to be non- or mildly-relativistic,
consisting of electrons and protons, and not radiating. Emission
comes from the relativistic ``clouds'' of electron-positron plasma
moving along the helical path wrapped around the jet and driven by
helical magnetic field. Brightness distribution on the VLBI map,
dependencies of the component parameters on the distance from the
core, as well as synchrotron spectra for certain geometries, were
presented in \citet{despringre1997}.

The helical jet model \citep[e.g.,][]{gomez1994a, villata1994}
presents the jet as an inhomogeneous flow of plasma along the
helical magnetic field filaments, interacting with the ambient
matter. It emits constantly, and flux density variability is
caused by the change in jet orientation. This model, if applied to
a number of sources including \object{0716+714}, explains its spectral
variability \citep{ostorero2001}.

\subsection{Long-term variation in the component speeds and ejection rate}
\citet{bach2005} find that the proper motions of the jet
components are almost monotonically slowing down during the period
from late 1986 until 1998 (0.86 to 0.29 mas/yr, respectively),
{\corr and \citet{nesci2005} explain this fact by a slow jet
precession. Their model predicts a further increase in the
apparent speed, together with decrease in the source's
mean brightness. Although} our data show with a certainty that
the speeds  in 2004 are faster than {\corr reported by
\citet{bach2005}, and in the period from 2004 to 2008, the source
is even slightly brighter in optical than in 1995 - 2003. It
indicates that jet precession is not the only mechanism
responsible for the changes in the apparent speed.} Also the
ejection interval has reduced dramatically: if each Gaussian
component in our model corresponds to a distinct physical feature
in the jet, the ejection rate after the outburst in 2004 is one in
0.17 years, which is much higher than the long-term mean of one in
1.2 years.

The deviations from the mean ejection rate during the period 1986 - 1998 have
occurred after small 15 GHz flux density enhancements. One explanation for this
behaviour could be that the mechanism that normally produces a
steady rate of ejections is disturbed during an outburst and
settles down gradually after that.

\section{Conclusions}
We studied the kinematics of the inner jet of \object{0716+714} at 43~GHz. Even
  though the jet has a very smooth brightness distribution, we
identified four components: three moving and one stationary.
Components are well-defined at several frequencies and traceable
over time. Components move along the jet with superluminal speeds
ranging from 8.5\,$c$ to 19.4\,$c$, which contradicts the model of
an oscillating jet.

Ejections of new superluminal components occur at a relatively
high rate, once in two months, and a component life time is
approximately ten months. Therefore, future observers have to note
that for accurate structure evolution studies, \object{0716+714} has to be
observed at a rate of once in one or two months. Superluminal
component ejections coincide with outbursts in a millimetre
continuum.

We also noticed that components move along wiggling trajectories
in the jet, which might be a signature of their helical motion.
Careful quantitative tests of models that predict physically
helical component motion (e.g. two-fluid model and helical jet
model) should be performed in the future in order to understand
nature of the jet in this object.

\begin{acknowledgements}
  The VLBA is a facility of the National Radio Astronomy Observatory, operated
  by Associated Universities, Inc., under cooperative agreement with the
  U.S. National Science Foundation.  T. Savolainen is a research fellow of the
  Alexander von Humboldt Foundation. This work was supported in part by Academy of Finland grant 120516.  K.~Sokolovsky is supported by the the
  International Max Planck Research School (IMPRS) for Radio and Infrared
  Astronomy.  We thank T. Hovatta and A. L{\"a}hteenm{\"a}ki for access to
  the Mets{\"a}hovi Blazar Monitoring 37 GHz data and flare model parameters.
\end{acknowledgements}
\bibliographystyle{aa} 
\bibliography{1425} 
\end{document}